\documentclass{PoS}
\usepackage{xspace}
\usepackage{graphicx}
\usepackage{caption}
\usepackage[labelformat=parens]{subcaption}
\usepackage{amsmath}
\usepackage{tikz}
\title{PID performance of the ALICE-TOF detector in Run 2}
\ShortTitle{PID performance of the ALICE-TOF detector in Run 2}
\author{\speaker{Nicol\`o Jacazio}\thanks{For the ALICE Collaboration}\\
  University of Bologna and INFN\\
  E-mail: \email{nicolo.jacazio@bo.infn.it}}
\newcommand{\MeVc}{MeV/\textit{c}\xspace}
\newcommand{\GeVc}{GeV/\textit{c}\xspace}
\newcommand{\pt}{$p_{\rm T}$\xspace}
\newcommand{\TOFl}{Time-Of-Flight\xspace}
\newcommand{\TOFm}{\ensuremath{Time\operatorname{-}Of\operatorname{-}Flight}\xspace}
\newcommand{\TOF}{TOF\xspace}
\newcommand{\PIDl}{Particle IDentification\xspace}
\newcommand{\PID}{PID\xspace}
\newcommand{\TPCl}{Time Projection Chamber\xspace}
\newcommand{\TPC}{TPC\xspace}
\newcommand{\ITSl}{Inner Tracking System\xspace}
\newcommand{\ITS}{ITS\xspace}
\newcommand{\PbPb}{Pb--Pb\xspace}
\newcommand{\snnt}[1]{\ensuremath{\sqrt{s_{NN}} =}~#1~TeV\xspace}
\newcommand{\snn}{\ensuremath{\sqrt{s_{NN}} = }\xspace}
\newcommand{\subfig}[1]{figure~(\subref{#1})\xspace}
\newcommand{\fig}[1]{Fig.~\ref{#1}\xspace}
\newcommand{\eq}[1]{Eq.~\ref{#1}\xspace}
\newcommand{\hi}{heavy-ion\xspace}
\newcommand{\tevent}{\ensuremath{t_{event}}\xspace}
\newcommand{\tTOF}{\ensuremath{t_{TOF}}\xspace}
\newcommand{\colltime}{event collision time\xspace}

\abstract{
  In these proceedings we report on the status of the ALICE \TOFl (\TOF) detector.
  The running performance of the Run 1 (2009-2013) and Run 2 (2015-present) data taking campaigns are compared.
  The \PIDl (\PID) capabilities of the detector are presented and discussed in the light of the improved detector calibration that allowed to reach a timing resolution of 56 ps.
}

\FullConference{Sixth Annual Conference on Large Hadron Collider Physics (LHCP2018)\\
  4-9 June 2018\\
  Bologna, Italy}

\begin{document}
  \section{Introduction}
  The ALICE experiment \cite{Aamodt:2008zz, Abelev:2014ffa} is one of the four main experiments recording collisions at the CERN Large Hadron Collider.
  One main point of its physics program is the study of the dense and hot matter created in ultra-relativistic \hi collisions i.e. the so called Quark-Gluon-Plasma.
  The ALICE experiment is not limited to the study of \hi collisions but it extends to smaller systems such as pp and p--Pb collisions, these cover interesting physics cases on themselves but are also be used as a useful reference for the \hi case.
  Identifying the particles produced in the collisions is fundamental in many analyses.
  This is the reason why the ALICE detector is capable of providing \PIDl (\PID) over a continuous interval of momentum \cite{Adam:2016acv}.
  The ALICE detector was designed to have good tracking capabilities down to very low \pt ($\sim 100$ \MeVc).
  In order to achieve this performance it uses a moderate magnetic field in the mid-rapidity region (0.2 or 0.5 T) coupled with a very low material budget.
  The high density of charged particles created in a single \hi collision requires high granularity detectors.
  The tracking is performed in the region close to the collision vertex with a Silicon detector (\ITSl or \ITS) and then continued in a \TPCl (\TPC) towards the outer layers of the detector.
  Both detectors provide an analogic measurement of the particle energy loss and this information allows to differentiate between different particle species for what concerns low momenta (below $\sim 800$ \MeVc).
  The \TOFl (\TOF) \cite{Dellacasa:2000kh, Cortese:2002kf} is used to complement these detectors and extend the momentum reach of the \PID for charged particles.
  \section{The ALICE \TOFl detector and its PID capabilities}
\begin{figure}
    \centering
    \hspace*{-1.6cm}
    \begin{subfigure}[t]{0.5\textwidth}
      \centering
      \includegraphics[trim={.7cm 0cm 3.cm 2.cm}, clip, height=5.45cm]{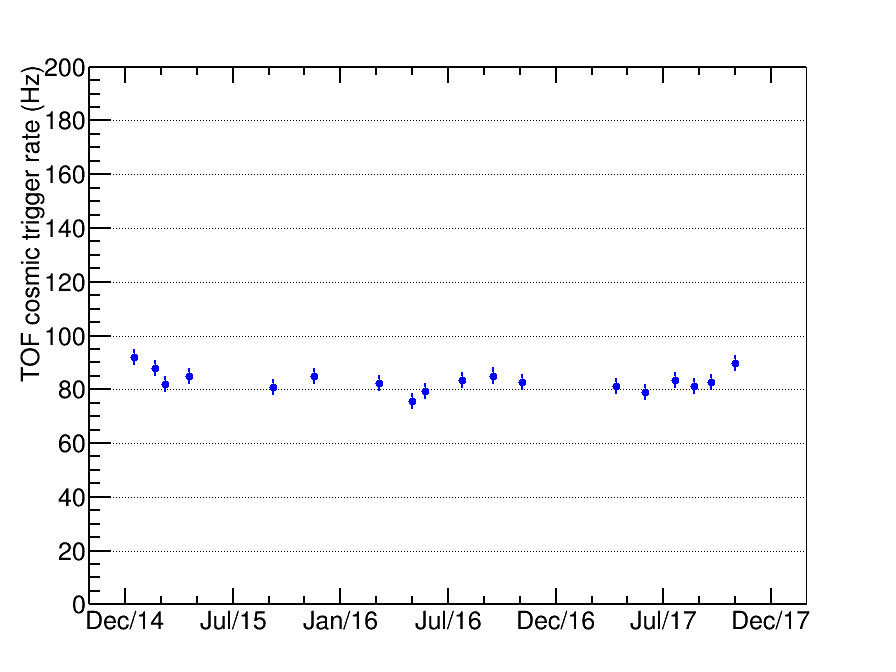}
      \caption{}
      \label{tofrate}
    \end{subfigure}%
    ~
    \hspace*{-1.25cm}
    \begin{subfigure}[t]{0.5\textwidth}
      \centering
      \begin{picture}(170,170)
        \put(0,0){\includegraphics[trim={1.4cm .23cm 1.8cm .92cm}, clip, height=5.6cm]{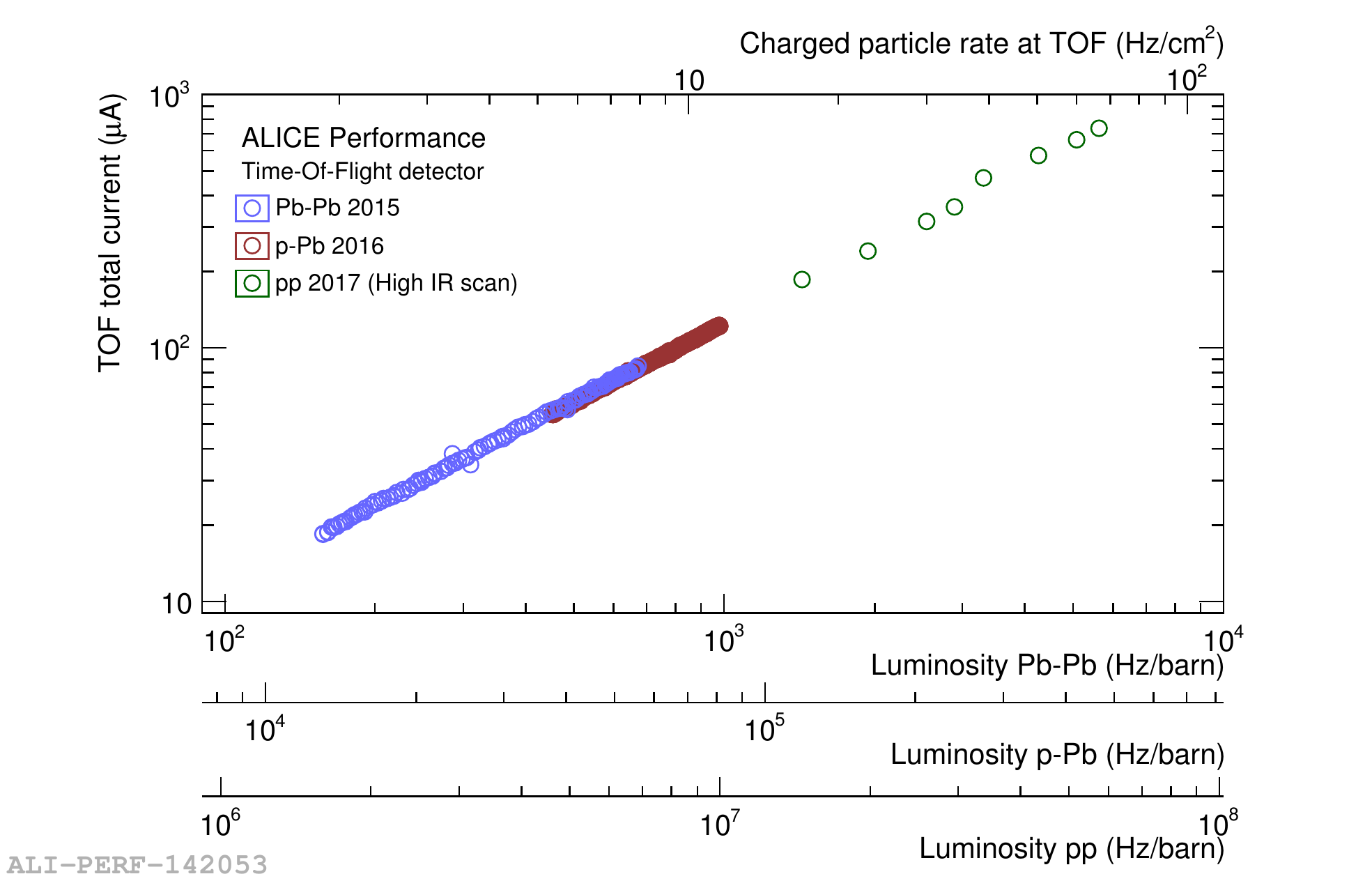}}
        \put(-2.8,-.5){  \tikz \fill [white] (0,0.) rectangle (1.2,.2); }
      \end{picture}
      \caption{}
      \label{currentlumi}
    \end{subfigure}%
    \\
    \begin{subfigure}[t]{0.5\textwidth}
      \centering
      \hspace*{-1.5cm}
      \includegraphics[trim={0.1cm 0 .2cm 0}, clip, height=6cm]{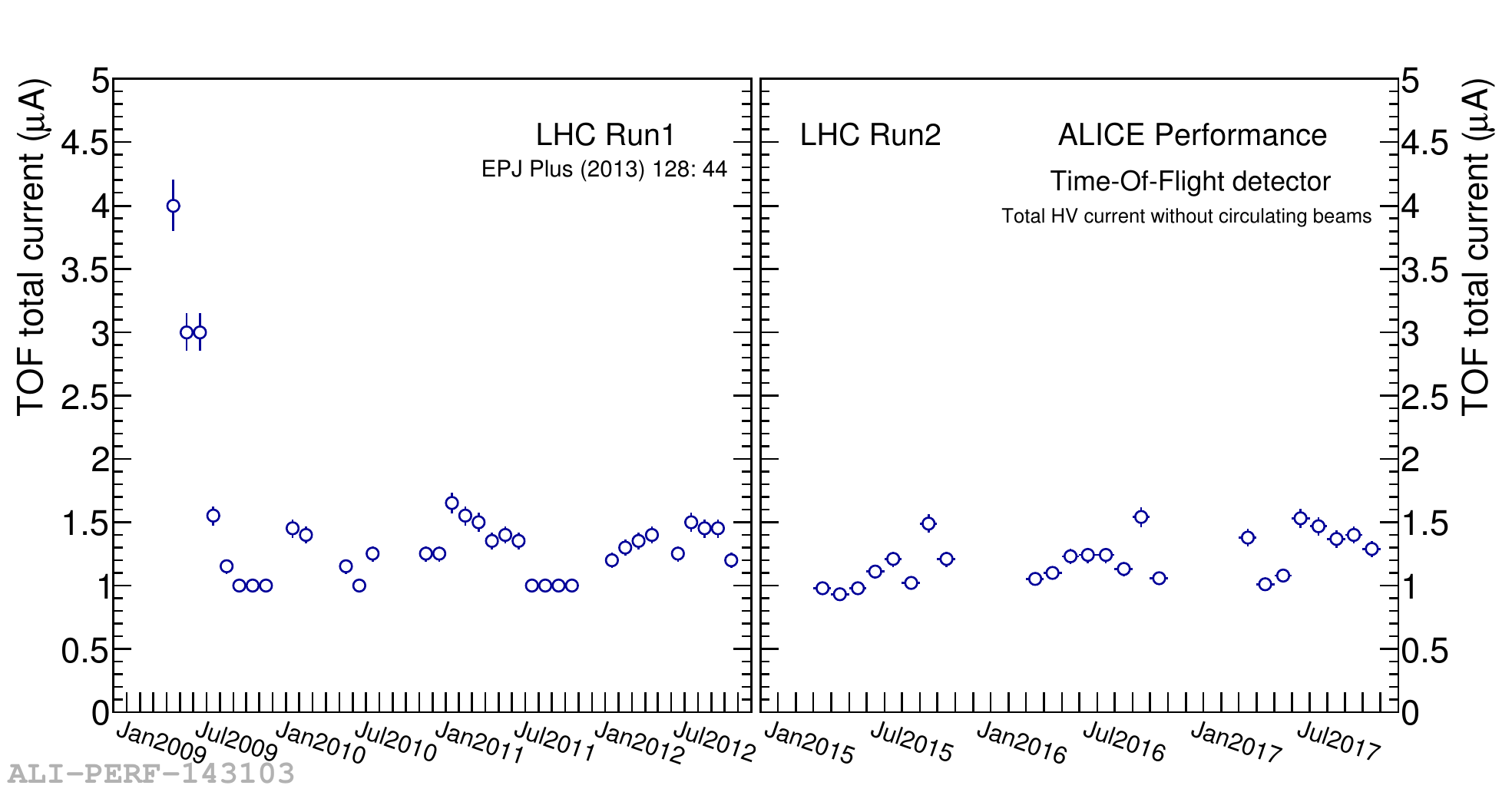}
      \caption{}
      \label{currentnobeam}
    \end{subfigure}%
    \caption{Summary of the running conditions of TOF during Run 1 and Run 2.
      In \subfig{tofrate} the rate of the cosmic events triggered by the TOF detector is shown as a function of time.
      In \subfig{currentlumi} and \subfig{currentnobeam} the total current passing trough the TOF MRPCs is shown as a function of the instantaneous luminosity (with beam) and as a function of time (without beam) respectively. 
    }
    \label{fig:operations}
  \end{figure}

\begin{figure}
    \includegraphics[trim={.1cm 0 1.65cm 1cm}, clip, height=6cm]{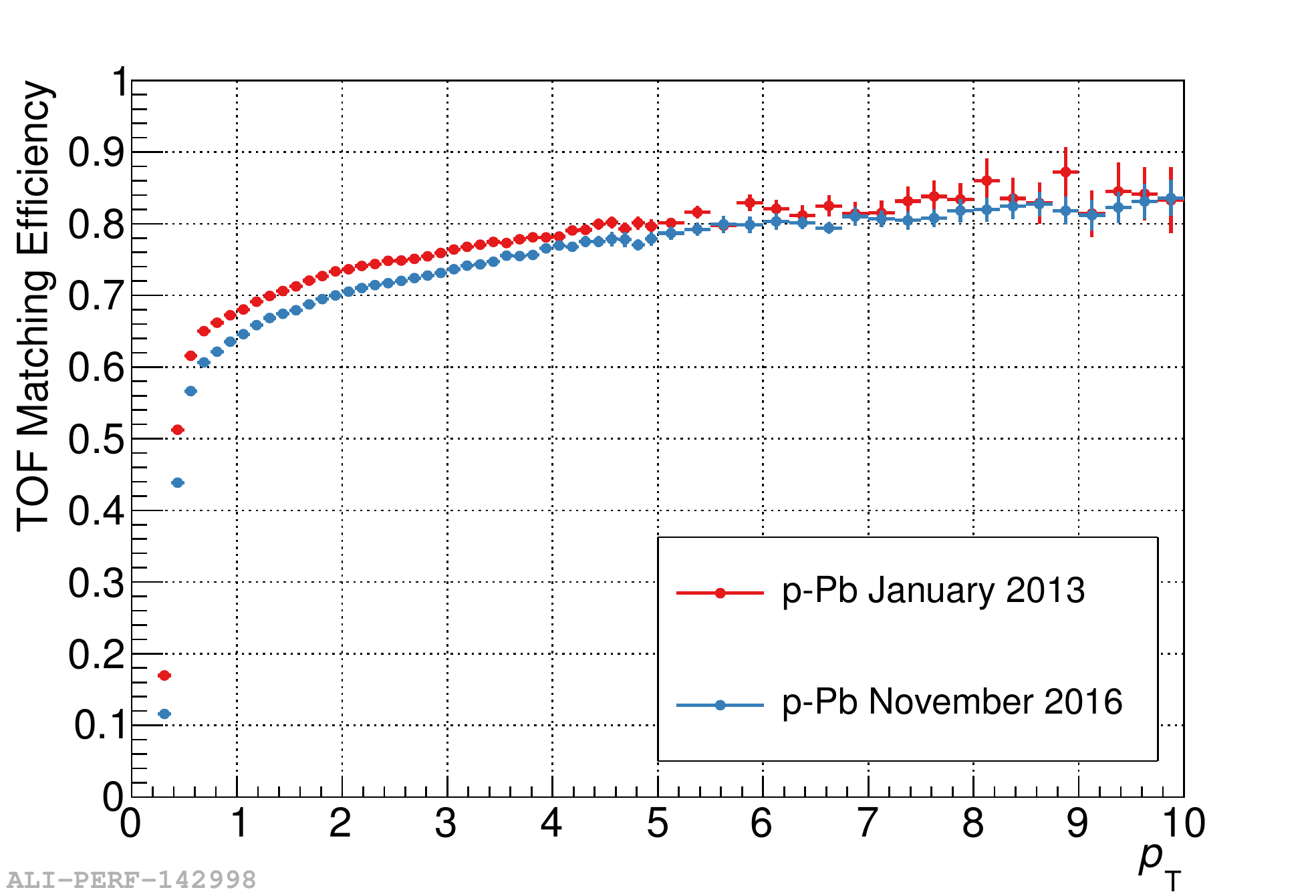}
    \centering
    \caption{
      Overall track matching efficiency as measured in p--Pb at $ \snn 5.02 $ TeV collisions during Run 1 and Run 2.
      The difference of the two efficiencies is due to the installation of all TRD modules.
    }
    \label{fig:efficiency}
  \end{figure}

\begin{figure}
    \centering
    \begin{subfigure}[t]{0.5\textwidth}
      \centering
      \begin{picture}(170,170)
        \put(0,0){\includegraphics[height=6cm]{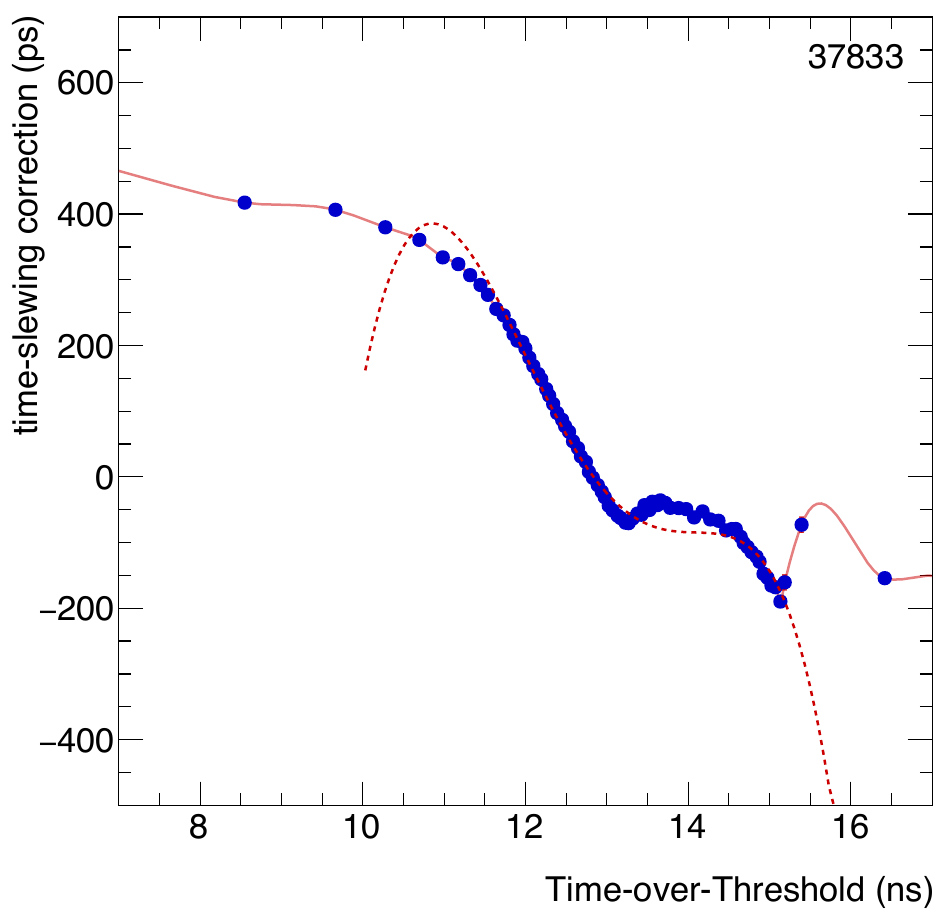}}
        \put(118,156){\tiny \TOF Ch. \#}
      \end{picture}
      \caption{}
      \label{singlechannel}
    \end{subfigure}%
    ~
    \begin{subfigure}[t]{0.5\textwidth}
      \centering
      \hspace*{-.8cm}
      \begin{picture}(170,170)
        \put(0,0){\includegraphics[height=6cm]{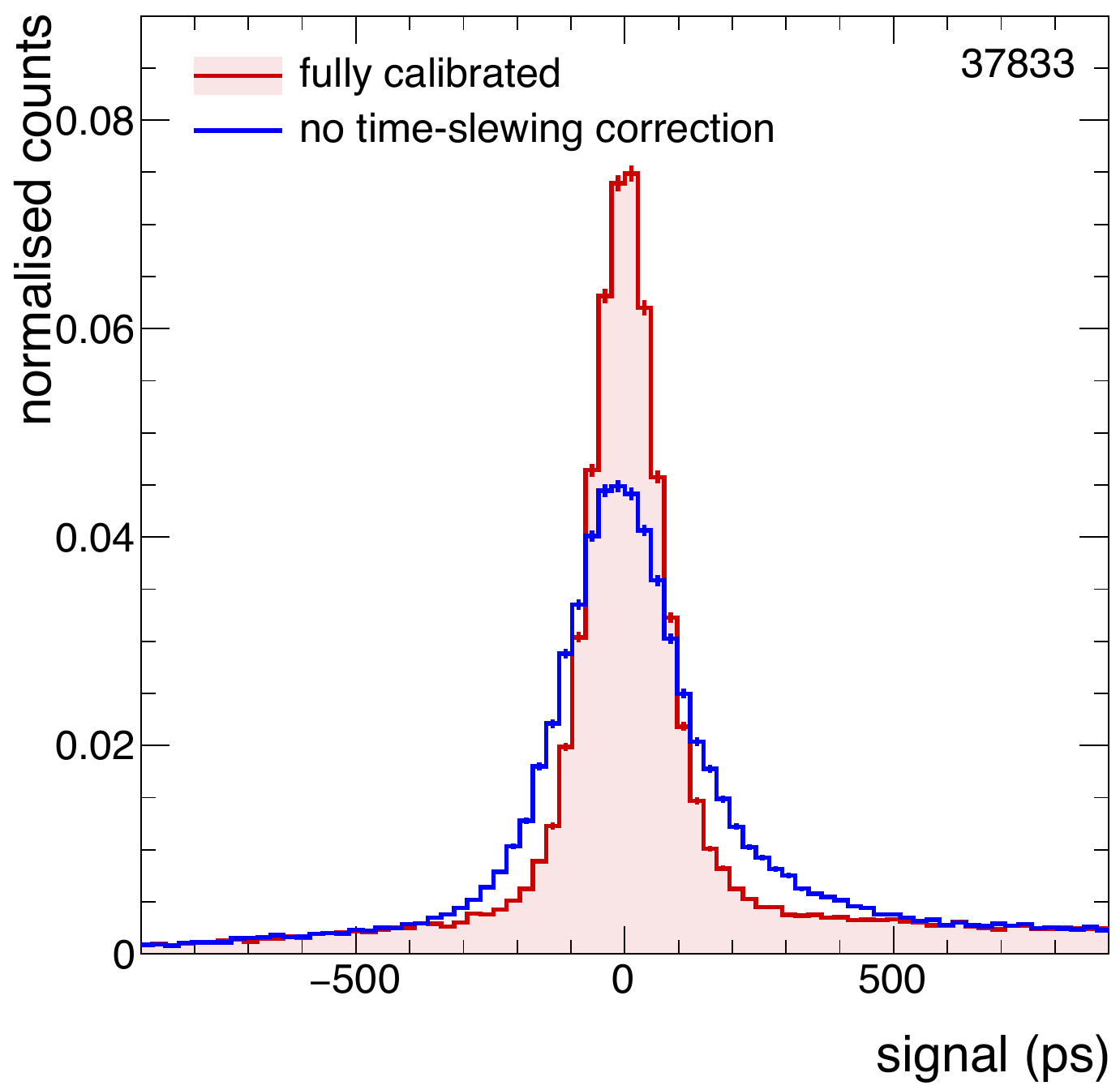}}
        \put(118,156){\tiny \TOF Ch. \#}
      \end{picture}
      \caption{}
      \label{calibration}
    \end{subfigure}%
    \caption{
      Summary of the status for the improved calibration procedure.
      In \subfig{singlechannel} the old (dashed line) and new (solid line) parameterization for the time slewing correction are drawn.
      In \subfig{calibration} the time signal distribution for a particular channel is shown before and after the time slewing correction is applied, it is clear that the correction improves the signal significantly.
    }
    \label{fig:calibration}
  \end{figure}

\begin{figure}
    \centering
    \begin{subfigure}[t]{0.5\textwidth}
      \centering
      \includegraphics[trim={0 0 .8cm .9cm}, clip, height=6cm]{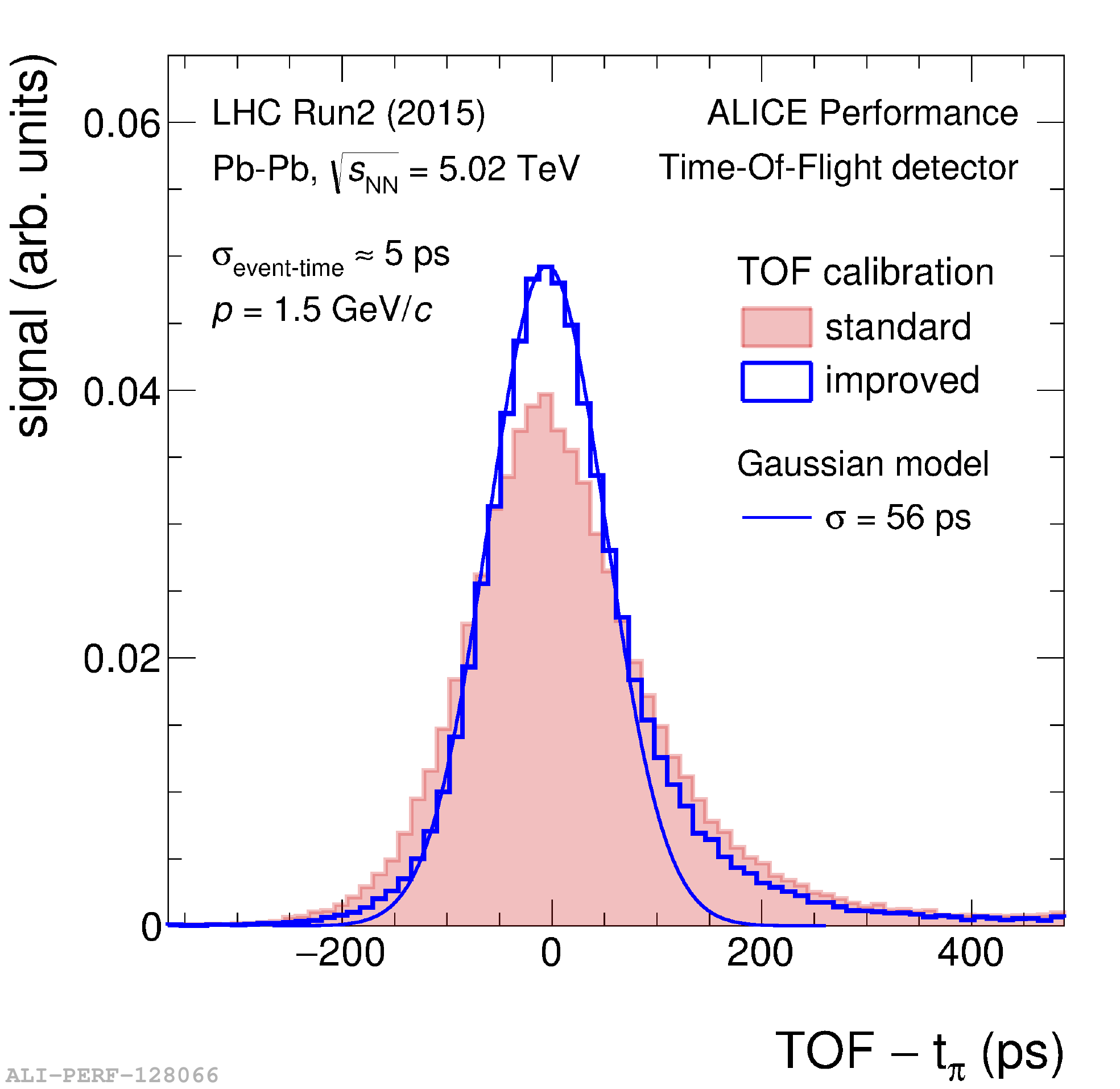}
      \caption{}
      \label{Run2resolution}
    \end{subfigure}%
    ~
    \begin{subfigure}[t]{0.5\textwidth}
      \centering
      \includegraphics[trim={0 0 .2cm 0cm}, clip, height=6cm]{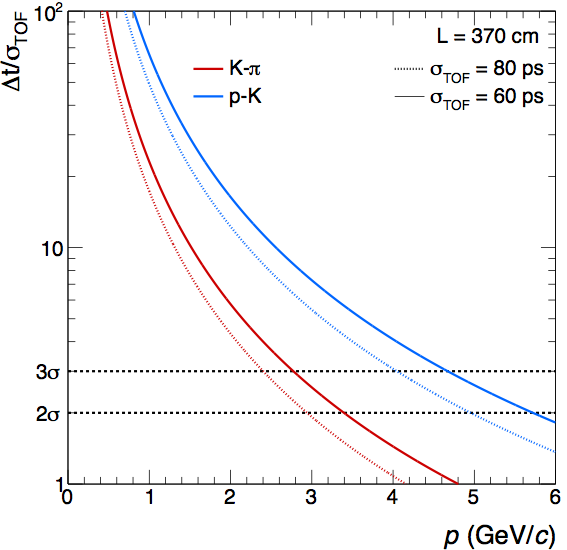}
      \caption{}
      \label{separation}
    \end{subfigure}%
    \caption{
      \subfig{Run2resolution} Signal distribution with resolution as measured by the full ALICE \TOF detector during Run 2.
      \subfig{separation} the impact of the improved resolution on the PID is depicted together with lines at 2 and 3 sigma. 
    }
    \label{fig:separation}
  \end{figure}

\begin{figure}
    \centering
    \begin{subfigure}[t]{0.5\textwidth}
      \centering
      \includegraphics[height=5.5cm]{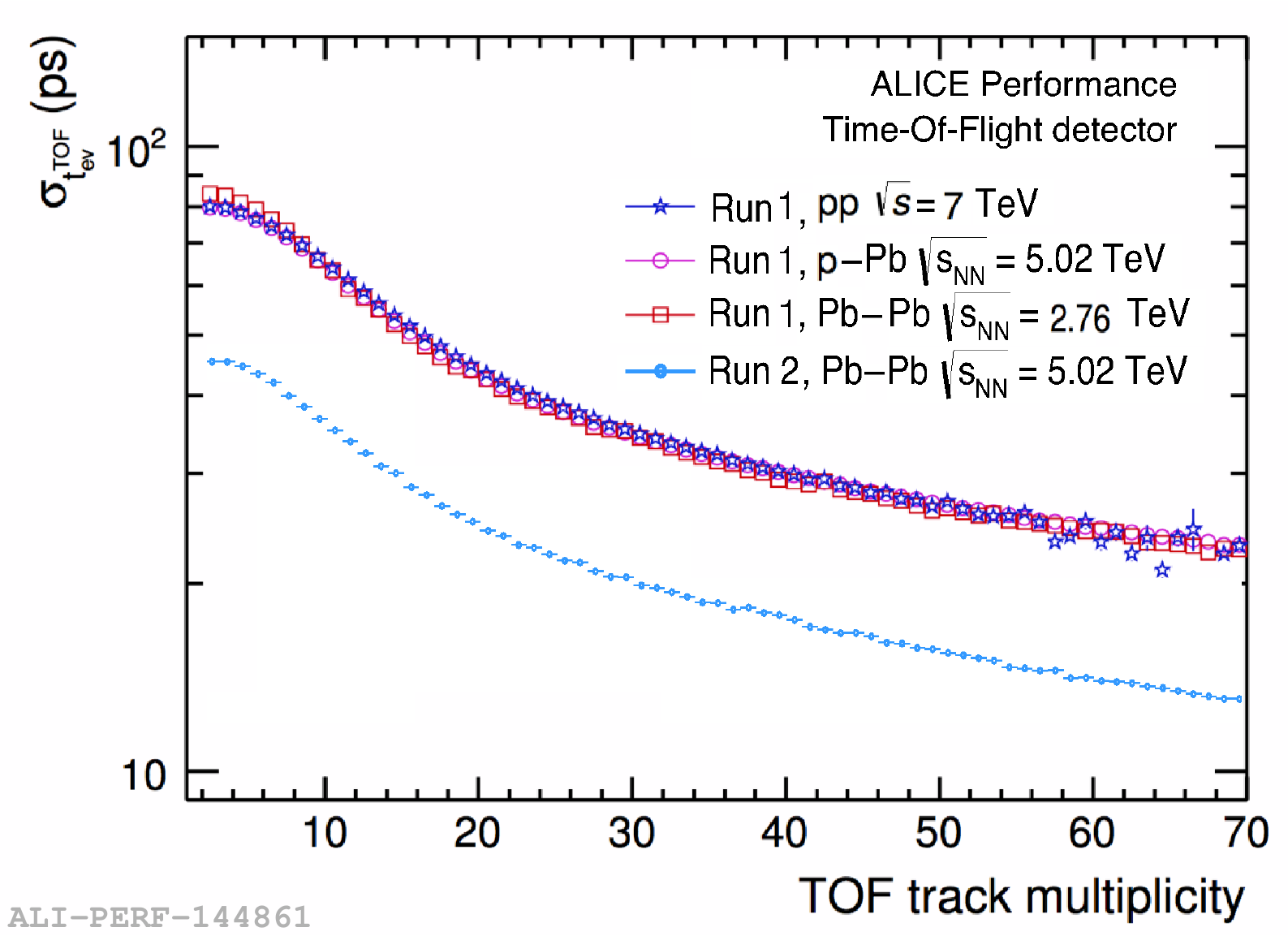}
      \caption{}
      \label{Run1ET}
    \end{subfigure}%
    ~ 
    \begin{subfigure}[t]{0.5\textwidth}
      \centering
      \vspace*{-5.5cm}
      \includegraphics[trim={1.3cm .8cm 0 .4cm}, clip, height=5.1cm]{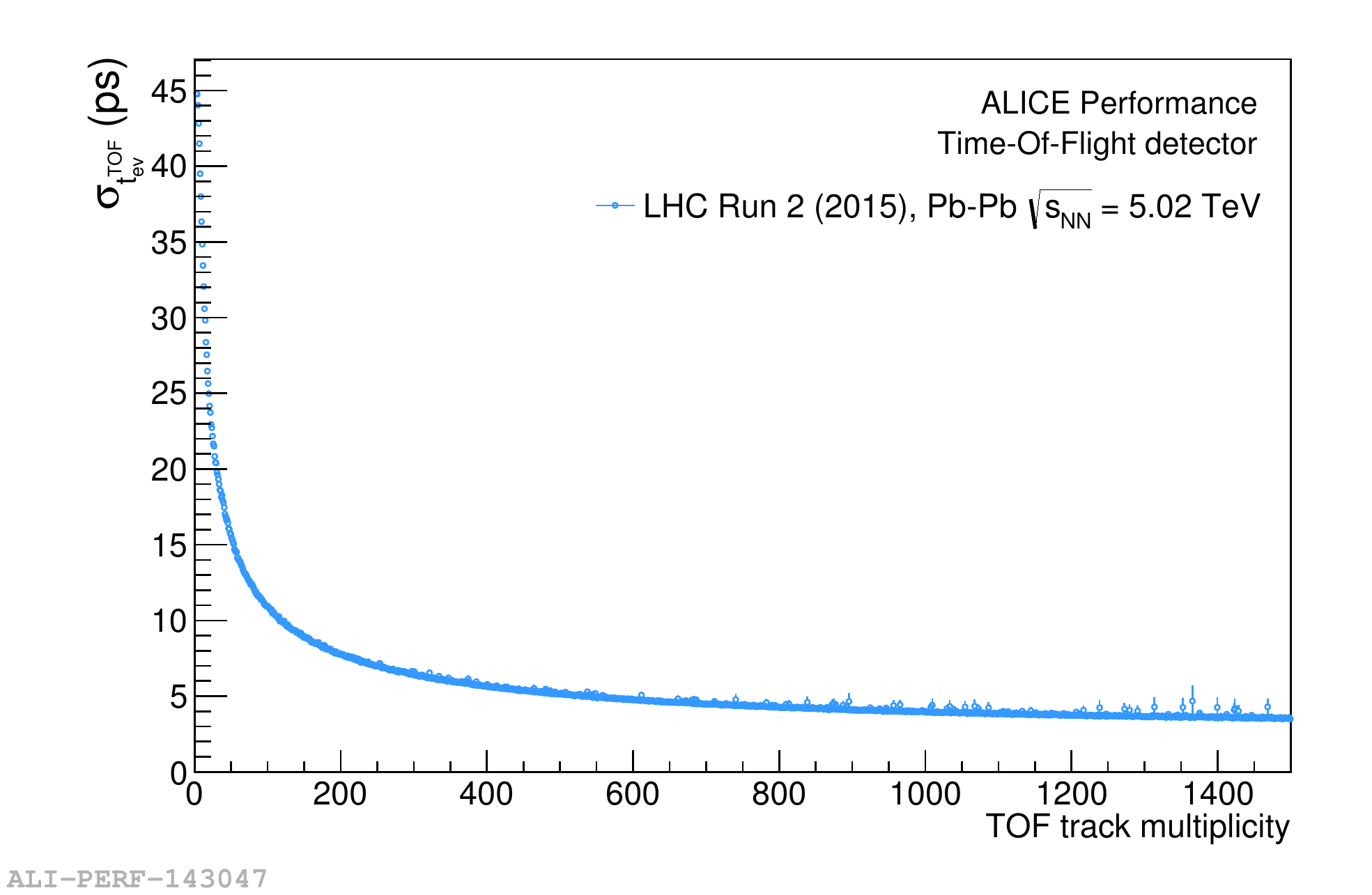}
      \caption{}
      \label{Run2ET}
    \end{subfigure}
    \caption{
      Collision \colltime resolution as measured with the \TOF detector in both Run 1 and Run 2 as a function of the number of reconstructed tracks with a time measurement.
      Two ranges are proposed: in \subfig{Run1ET} the \colltime resolution is shown up to multiplicities of 70 and compared across multiple collision systems, in \subfig{Run2ET} the multiplicity range is enlarged so as to fit the values reached in \hi collisions.
    }
    \label{eventtime}
  \end{figure}

\begin{figure}
    \includegraphics[trim={6.2cm 1.8cm 4.8cm 3.6cm}, clip, height=5.9cm]{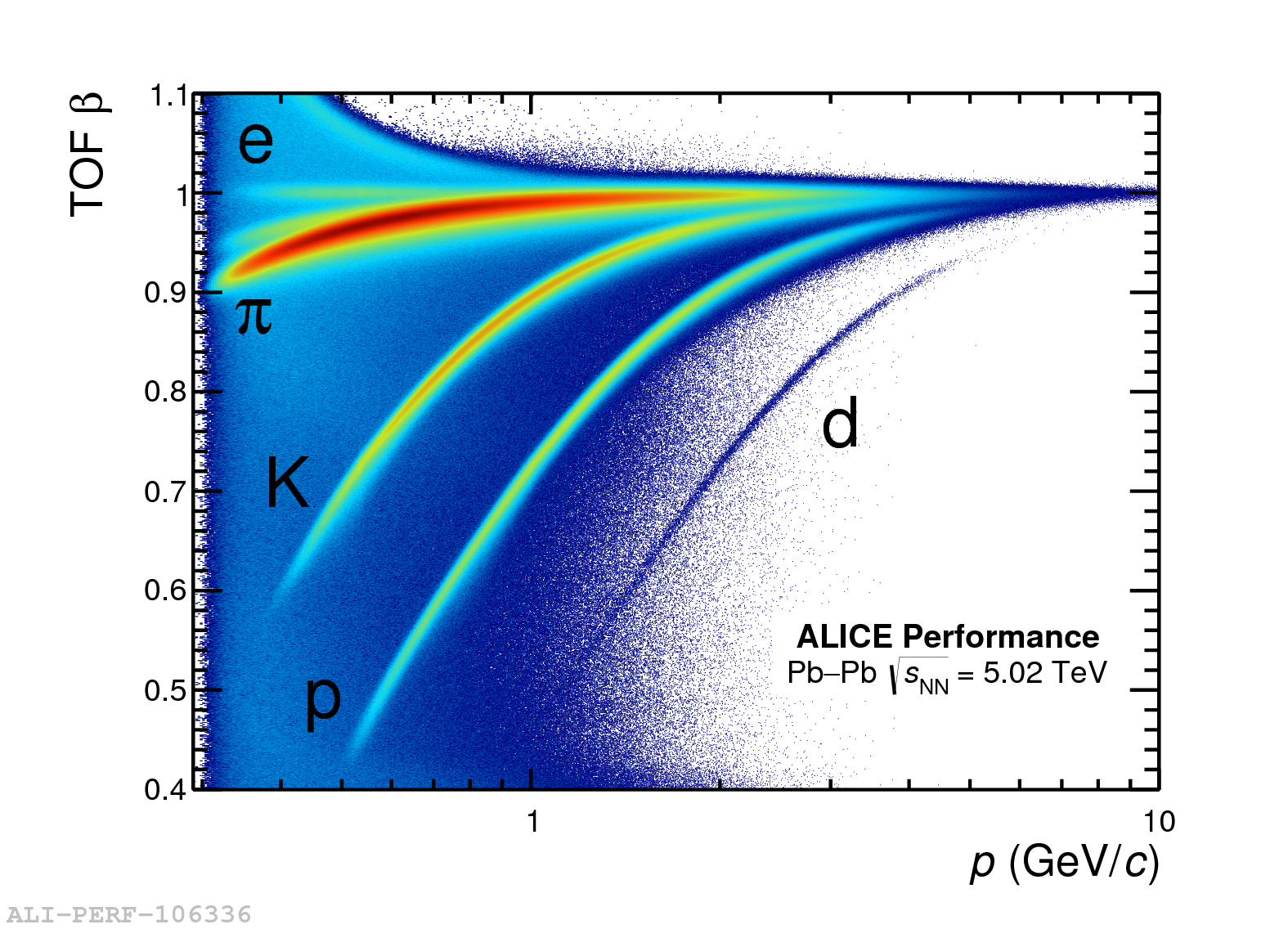}
    \centering
    \caption{Particle velocity ($\beta$) distribution as a function of momentum for tracks reconstructed with the ALICE experiment in \PbPb collisions at \snn 5.02 TeV.}
    \label{fig:betap}
  \end{figure}

\begin{figure}
    \centering
    \begin{subfigure}[t]{0.5\textwidth}
      \centering
      \includegraphics[trim={.1cm 0cm 1.5cm 1.15cm}, clip, height=4.6cm]{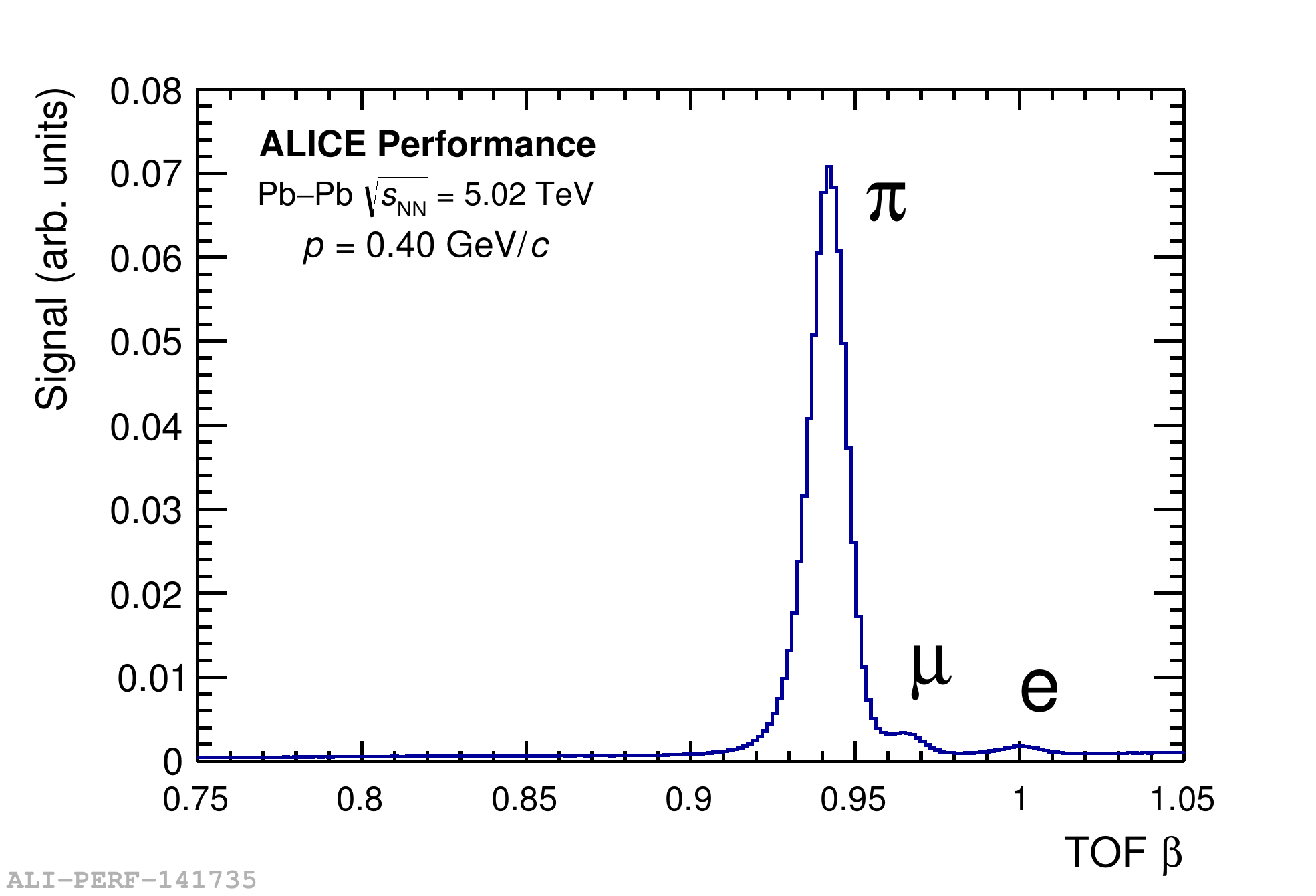}
      \caption{}
      \label{mev400}
    \end{subfigure}%
    ~
    \begin{subfigure}[t]{0.5\textwidth}
      \centering
      \includegraphics[trim={.1cm 0cm 1.5cm 1.15cm}, clip, height=4.6cm]{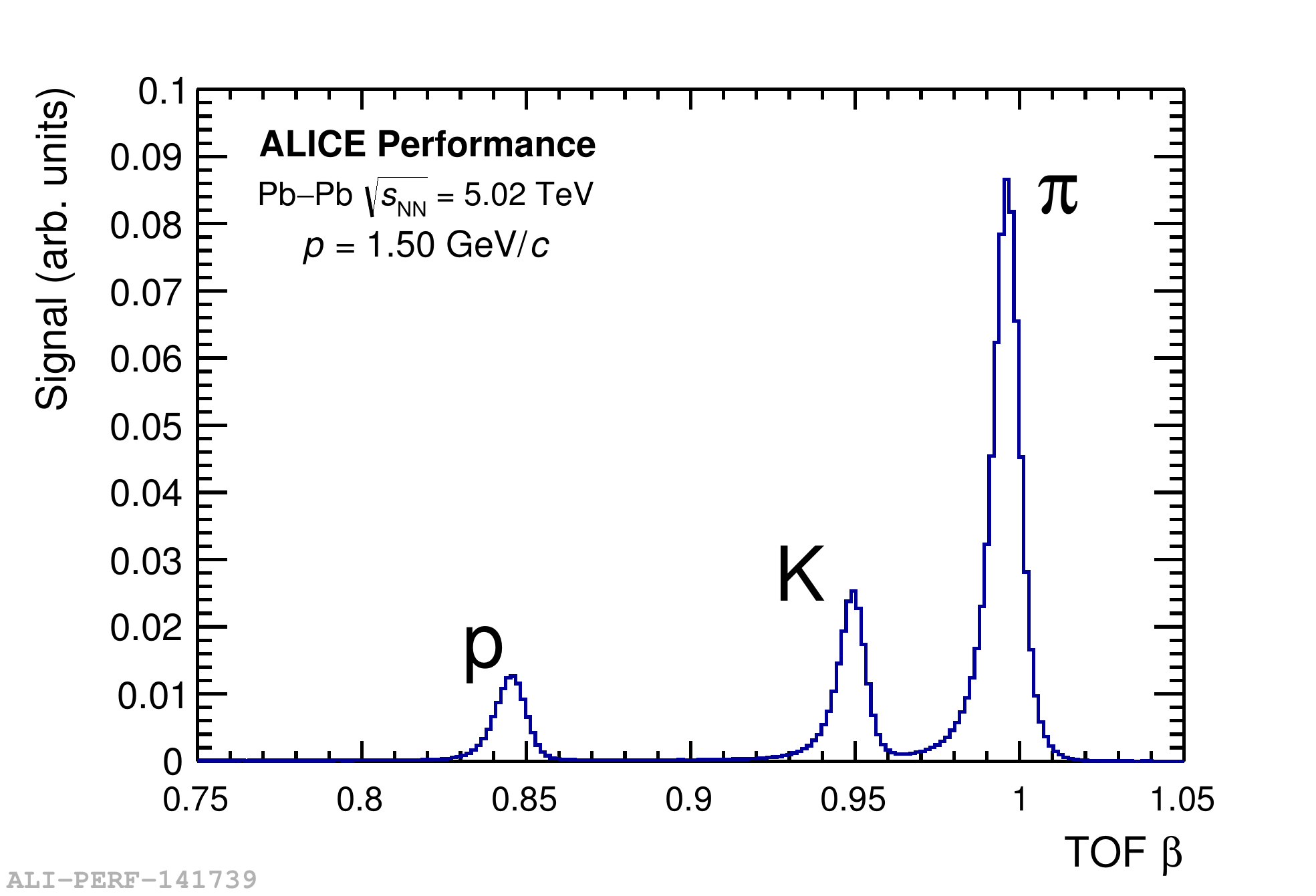}
      \caption{}
      \label{mev1500}
    \end{subfigure}%
    \\
    \begin{subfigure}[t]{0.5\textwidth}
      \centering
      \includegraphics[trim={.1cm 0cm 1.5cm 1.15cm}, clip, height=4.6cm]{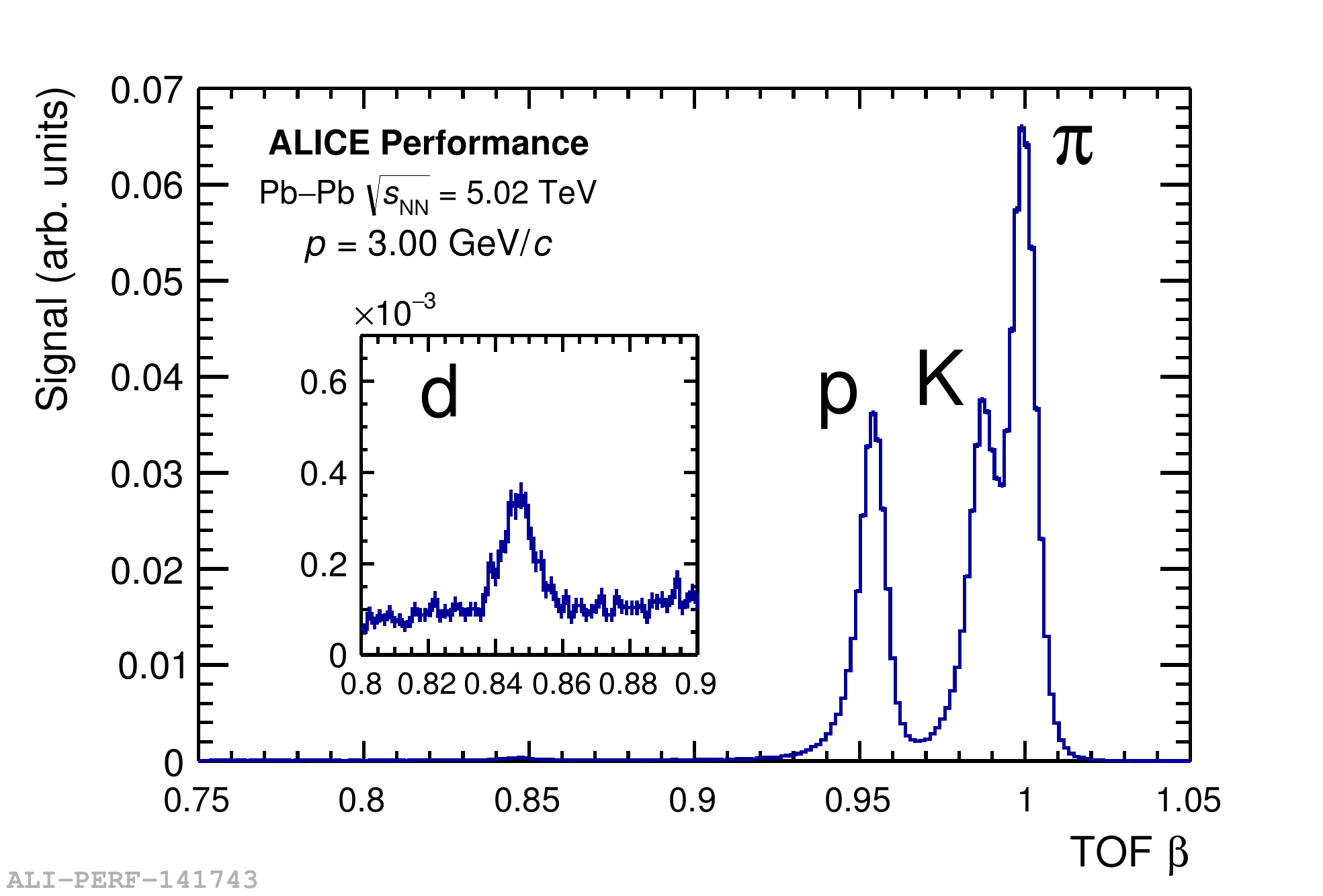}
      \caption{}
      \label{mev3000}
    \end{subfigure}%
    ~
    \begin{subfigure}[t]{0.5\textwidth}
      \centering
      \includegraphics[trim={.1cm 0cm 1.5cm 1.15cm}, clip, height=4.6cm]{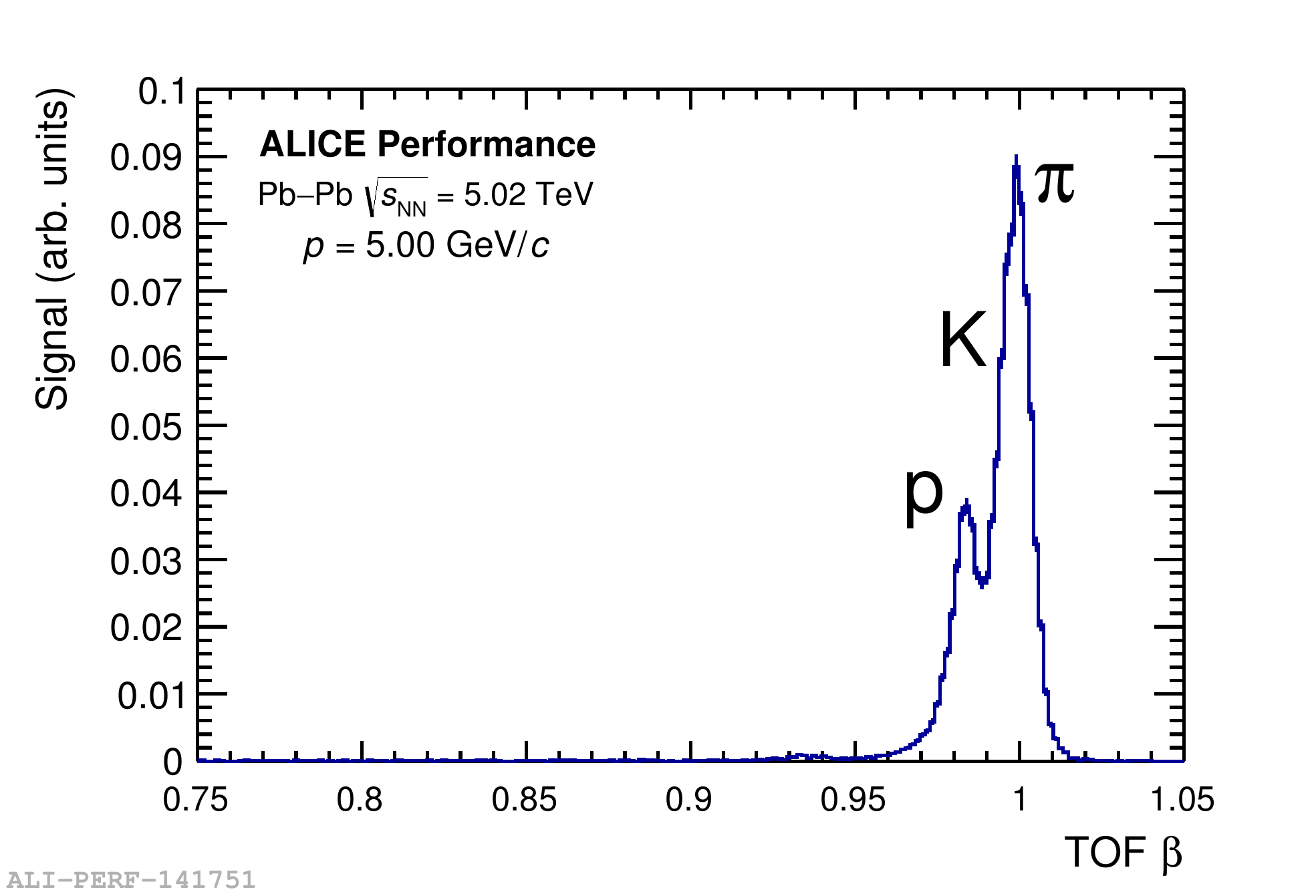}
      \caption{}
      \label{mev5000}
    \end{subfigure}%
    \caption{Distribution of the particle velocity ($\beta$) as measured with the \TOF detector in different momentum intervals.
      The signals from each particle species are clearly visible up to high momenta.
    }
    \label{fig:PIDcuts}
  \end{figure}

  The \TOF detector covers a large area ($\sim 141 {\rm m}^{2}$) with cylindrical geometry with internal radius $\sim 3.7$ m from the beam axis.
  With respect to the interaction point the TOF system uniformly covers a pseudorapidity ($\eta$) region of 0.9 with axial symmetry, for a total coverage of 1.8 units of $ \eta $ in the full azimuthal angle.
  The sensitive unit of the TOF detector is based on the Multigap Resistive Plate Chamber (MRPC) technology, built from two stacks of five single gas gaps. 
  The whole detector is segmented into 1593 MRPC strips, each one covering  $120 \times 7.4\ {\rm cm}^{2}$ and mounted inside 87 separate modules which are grouped into 18 sectors covering the whole $2\pi$ angle.
  To achieve the high granularity needed to guarantee a low occupancy even at the high charged particle densities of \PbPb collisions, the single MRPC strip is segmented into two rows of 48 pickup pads of $3.5 \times 2.5\ {\rm cm}^{2}$, accounting for 96 readout pads per strip and 152928 total readout channels.
  In addition to its \PID capabilities, the \TOF system provides a dedicated trigger for cosmic rays.
  In all its years of operation the \TOF detector proved itself to be very reliable: an example can be seen in \fig{tofrate} where the trigger rate for cosmic events is shown as a function of time and it is found to be stable, indicating no sign of detector degradation so far.
  The same conclusion can be drawn from the measurement of the current passing trough all 1593 MRPCs.
  In \fig{currentnobeam} the current without any beam circulating in the LHC is plotted as a function of time and it is found to be very stable with the exception of few points measured just after the finalization of the detector installation and due to the initial MRPCs HV conditioning.
  The stability of the current and its low value indicate that the detector has low noise and no signs of ageing is observed from 2009 till 2017.
  The current is also shown as a function of the instantaneous luminosity in \fig{currentlumi}, the stability and the linearity of the response without any sign of saturation make the TOF detector a good candidate for a luminometer.
  The ALICE \TOF MRPC was designed to ensure large particle detection efficiencies ($\sim 99\%$), nonetheless this efficiency has to be coupled with the detector geometrical acceptance and the global tracking algorithm of the ALICE experiment, resulting in a lower effective efficiency shown in \fig{fig:efficiency} for p--Pb collisions as measured from Run 1 and Run 2 data.
  The discrepancies between the two periods can be understood as follows: during the Long Shutdown 1 phase ALICE finished the installation of all the modules of the TRD detector which is located between TPC and TOF, thus adding more material for particles to cross. 
  These changes are the cause for the difference in the overall track matching efficiencies, indicating again no sign of ageing of the MRPCs.
  The data collected in the first months after the start of Run 2 was used to extract an improved calibration.
  Among the several technical changes used for the calibration it is important to mention the single channel granularity (in the past 8-channel groups were used) and the use of a finer point by point parameterization of the correction as a function of the measured Time-Over-Threshold (in the past limited to a $5^{th}$ order polynomial).
  Both standard and improved calibration curves are shown for one channel in \fig{singlechannel}.
  The effect of the calibration on a single channels is shown in \fig{calibration} and during the whole 2017 the new procedure proved to be very stable as 88\% of all TOF channels showed RMS lower than 20 ps.
  The \TOFm measurement can simply be obtained by subtracting the \colltime (\tevent) from the time measured with the \TOF detector (\tTOF), as reported in \eq{eq:1}:
  \begin{align}
    \begin{split}\label{eq:1}
      \TOFm \equiv TOF = \tTOF - \tevent
    \end{split}
    \\
    \begin{split}\label{eq:2}
      \sigma_{TOF} &= \sigma_{ \tTOF}^{2}+ \sigma_{tracking}^{2} +\sigma_{\tevent}^{2} \\
      \sigma^{2}_{\tTOF} &= \sigma^{2}_{MRPC} + \sigma^{2}_{TDC} +\sigma^{2}_{FEE} +\sigma^{2}_{Calibration}
    \end{split}
    \\
    \begin{split}\label{eq:3}
      \sigma_{\tevent} &\sim \frac{\sigma_{\tTOF} }{ \sqrt{n_{\rm tracks}}} 
    \end{split}
  \end{align}
  The components that contribute to the uncertainty on \tTOF and \tevent are listed in \eq{eq:2} and \eq{eq:3}.
  The resolution on the simple time measurement ($\sigma_{\tTOF}$) depends mostly on fixed elements which are intrinsic of the detector such as the MRPC resolution ($\sigma_{MRPC}$) and the one given by electronic readout chain ( $\sigma_{TDC}$ and $\sigma_{FEE}$).
  Thanks to the improvements described above the contribution due to the calibration ($\sigma_{Calibration}$) was significantly reduced and allowed to achieve resolution of $\sim 56 ps$, as reported in \fig{Run2resolution}.
  The overall effect on the separation power is depicted in \fig{separation}.
  The determination of the \colltime needed to compute the $\TOFm$ can be performed autonomously by the \TOF detector itself with the iterative procedure described in \cite{Adam:2016ilk}.
  Thanks to this method the measurement of the $\TOFm$ can benefit further from the improved precision on \tTOF as reported in \eq{eq:3}.
  The $\sigma_{\tevent}$ depends on the number of tracks available for the \tevent evaluation ($n_{\rm tracks}$).
  The improvement between Run 1 and Run 2 of the \colltime precision is shown in \fig{Run1ET}.
  With the improved calibration the resolution is found to be better than 5 ps for events with more than $\sim 600$ tracks matched to \TOF (\fig{Run2ET}).
  All the developments and procedures described above are summarized in \fig{fig:betap} where the $\beta$ of each particle is shown as a function of its momentum as measured in \PbPb collision at \snnt{5.02}.
  The momentum slices of the $\beta$ distribution are shown in \fig{fig:PIDcuts}, where it is possible to appreciate the separation of the signals of each particle species.
  Starting at 400 \MeVc in \fig{mev400} it is possible to identify the signals of pions, muons and electrons.
  At larger momenta (\fig{mev1500}) the peaks of protons and kaons are visible, while light nuclei can also be identified starting from $\sim 3$ \GeVc (\fig{mev3000}).
  As expected from \fig{separation} the proton and kaon peaks are still separated at 5 \GeVc (\fig{mev5000}).
  \section{Conclusions}
  The working performance of the ALICE \TOFl detector was discussed and it was shown that the detector does not appear to suffer from ageing in any way.
  This information is crucial as the \TOF detector will continue the data taking also after Run 2 with limited modifications to the readout chain.
  During Run 2 an improvement in the TOF calibration made it possible to reach timing resolutions of $\sim 56$ ps, thus improving significantly the \PIDl capabilities and extending its reach.
  In the light of these new developments the PID performance now allows an improved separation of electrons, muons, pions, kaons, protons and light nuclei over a wide range of momenta.
  \bibliography{bibliography}
  \bibliographystyle{JHEP}
\end{document}